\documentclass[twocolumn,twoside,slac_two]{revtex4}
\usepackage{graphicx}
\usepackage{fancyhdr}
\begin{document}

%Comment: Flavor Physics & CP Violation Conference, Bled, 2007
\title{Factorization Approaches to $B$ Meson Decays}


\author{Hsiang-nan Li}
\affiliation{Institute of Physics, Academia Sinica, Nankang,
Taipei 115, Taiwan, Republic of China} \affiliation{Department of
Physics, National Cheng-Kung University, Tainan 701, Taiwan,
Republic of China} \affiliation{Department of Physics, National
Tsinghua University, Hsinchu 300, Taiwan, Republic of China}

\begin{abstract}
We compare the theoretical frameworks and the phenomenological
applications of the factorization approaches to exclusive $B$
meson decays, which include QCD-improved factorization,
perturbative QCD, and soft-collinear effective theory. Recent
progress on two-body nonleptonic $B$ meson decays made in these
approaches are reviewed.
\end{abstract}

\maketitle

\thispagestyle{fancy}


\section{Introduction}

``Factorizations" in the naive factorization assumption and in
factorization theorem have very different meanings. The former
refers to the factorization of a process into subprocesses. For
example, a $B$ meson decay amplitude $A(B\to M_1M_2)$ is written,
in the factorization assumption, as the product \cite{BSW},
\begin{equation}
A(B\to M_1M_2)\propto f_{M_2}F^{BM_1}\;,\label{fa}
\end{equation}
where the meson decay constant $f_{M_2}$ arises from the
production of the meson $M_2$ from the vacuum, and the form factor
$F^{BM_1}$ is associated with the $B\to M_1$ transition. The
latter refers to the factorization of perturbative and
nonperturbative dynamics in a QCD process. According to
factorization theorem, the above amplitude is expressed as
\begin{eqnarray}
A(B\to M_1M_2)\propto \phi_B\otimes H\otimes
\phi_{M_1}\otimes\phi_{M_2}\;,\label{ft}
\end{eqnarray}
where $\otimes$ denotes the convolution over parton kinematic
variables, the hard kernel $H$ absorbs perturbative dynamics, and
the $B$ ($M_1$, $M_2$) meson distribution amplitude $\phi_B$
($\phi_{M1}, \phi_{M2}$) absorbs nonperturbative dynamics in the
$B\to M_1M_2$ decay. A piece of contribution to $B$ meson decays
is factorizable, if it respects either Eq.~(\ref{fa}) in the sense
of the factorization assumption, or Eq.~(\ref{ft}) in the sense of
factorization theorem. Below we shall use the terms
''factorizable" and ''nonfactorizable" without specifying which
sense it refers to.

\section{Collinear vs. $k_T$ Factorization}

Both collinear and $k_T$ factorizations are the fundamental tools
of perturbative QCD, where $k_T$ denotes parton transverse
momenta. We first explain these two types of theorems by
considering the simplest scattering process
$\pi(P_1)\gamma^*\to\gamma(P_2,\epsilon)$ as an example. The
momentum $P_1$ of the pion and the momentum $P_2$ of the out-going
on-shell photon are chosen as
\begin{eqnarray}
P_1 =(P_1^+,0,{\bf 0}_T)\;, \;\;\; P_2 = (0,P_2^-,{\bf
0}_T)\;.\label{ppb}
\end{eqnarray}
The leading-order (LO) quark diagram, in which the anti-quark
$\bar q$ carries the on-shell fractional momentum
$k=(xP_1^+,0,{\bf 0}_T)$ and the internal quark carries $P_2-k$,
leads to the amplitude,
\begin{eqnarray}
G^{(0)}(x,Q^2)&=&\frac{tr[\not \epsilon (\not P_2-\not k)
\gamma_\mu \not P_1\gamma_5]}{(P_2-k)^2}\nonumber\\
&=&-\frac{tr[\not \epsilon \not P_2 \gamma_\mu \not
P_1\gamma_5]}{x Q^2}\;, \label{p1a}
\end{eqnarray}
with the leading spin structure $\not P_1\gamma_5$ of the pion and
the momentum transfer squared $Q^2\equiv 2P_1\cdot P_2$. We have
suppressed other constant factors, such as the electric charge,
the color number, and the pion decay constant, which are
irrelevant in the following discussion.

The trivial collinear factorization of Eq.~(\ref{p1a}) reads,
\begin{eqnarray}
G^{(0)}(x,Q^2)&=&\int dx'\phi^{(0)}(x;x')
H^{(0)}(x',Q^2)\;,\nonumber\\
\phi^{(0)}(x;x')&=& \delta(x-x')\;,
\nonumber\\
H^{(0)}(x,Q^2)&=&- \frac{tr[\not \epsilon \not P_2 \gamma_\mu \not
P_1\gamma^5]}{x Q^2}\;. \label{h0c}
\end{eqnarray}
The zeroth-order distribution amplitude $\phi^{(0)}$ is
proportional to $\delta(x-x')$, implying that the parton entering
the LO hard kernel $H^{(0)}$ carries the same momentum as the
parton entering the distribution amplitude does. The trivial $k_T$
factorization of Eq.~(\ref{p1a}) reads \cite{NL2},
\begin{eqnarray}
G^{(0)}(x,Q^2)&=&\int dx'd^2k'_T\Phi^{(0)}(x;x',k'_T)\nonumber\\
& &\times H^{(0)}(x',Q^2,k'_T)\;,\nonumber\\
\Phi^{(0)}(x;x',k'_T)&=& \delta(x-x')\delta({\bf k}'_T)\;,
\nonumber\\
H^{(0)}(x,Q^2,k_T)&=&- \frac{tr[\not \epsilon \not P_2 \gamma_\mu
\not P_1\gamma^5]}{x Q^2+k_T^2}\;. \label{h0p}
\end{eqnarray}
Because of the zeroth-order wave function
$\Phi^{(0)}\propto\delta({\bf k}'_T)$, $H^{(0)}$ does not depend
on the parton transverse momentum actually.

\begin{figure}[h]
\centering
\includegraphics[width=80mm]{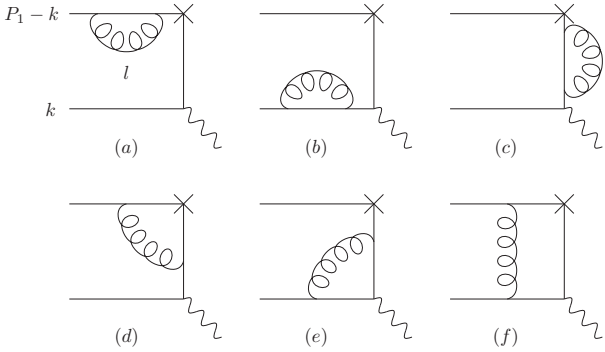}
\caption{$O(\alpha_s)$ quark diagrams for $\pi\gamma^*\to\gamma$
with $\times$ representing the virtual photon vertex.}
\label{fig1}
\end{figure}

\begin{figure}[h]
\centering
\includegraphics[width=80mm]{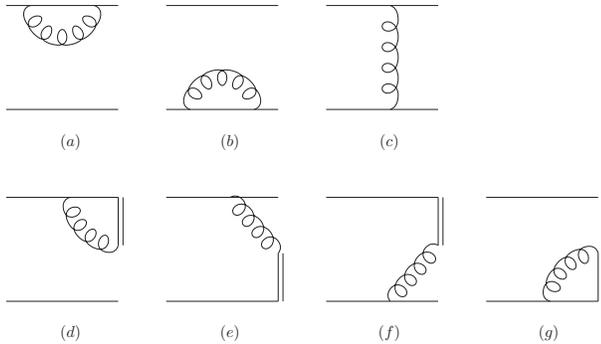}
\caption{$O(\alpha_s)$ effective diagrams for the pion wave
function.} \label{fig2}
\end{figure}

The $O(\alpha_s)$ quark diagrams for Eq.~(\ref{p1a}) from full QCD
are displayed in Fig.~\ref{fig1}, in which the upper line
represents the $q$ quark. The collinear factorization of these
radiative corrections is given by
\begin{eqnarray}
G^{(1)}(x,Q^2)&=&\int dx'
\phi^{(1)}(x;x')H^{(0)}(x',Q^2)\nonumber\\
& & +H^{(1)}(x',Q^2)\;, \label{pgc1}
\end{eqnarray}
where the first-order distribution amplitude $\phi^{(1)}$ is
defined by the effective diagrams in Fig.~\ref{fig2} \cite{L1}.
Expressions from Figs.~\ref{fig2}(c), \ref{fig2}(e), and
\ref{fig2}(f) are proportional to $\delta(x-x'-l^+/P_1^+)$, where
$l$ is the loop momentum carried by the collinear gluon. The
$\delta$-function indicates that the exchange of the collinear
gluon modifies the momentum fraction of the parton entering
$H^{(0)}$ from $x$ to $x-l^+/P_1^+$. The $k_T$ factorization of
Fig.~\ref{fig1} leads to \cite{NL2}
\begin{eqnarray}
G^{(1)}(x,Q^2)&=&\int dx' d^2k'_T
\Phi^{(1)}(x;x',k'_T)\label{pg11}\\
& &\times H^{(0)}(x',Q^2,k'_T)+H^{(1)}(x,Q^2)\;,\nonumber
\end{eqnarray}
where the first-order wave function $\Phi^{(1)}$ from
Figs.~\ref{fig2}(c), \ref{fig2}(e), and \ref{fig2}(f) is
proportional to $\delta(x-x'-l^+/P_1^+)\delta({\bf k}'_T+{\bf
l}_T)$. In this case the collinear gluon exchange modifies both
the longitudinal and transverse parton momenta flowing into the
hard kernel.

It is observed that $H^{(0)}$ in Eq.~(\ref{pg11}) depends on
$k'_T$ nontrivially in the first-order $k_T$ factorization. Being
convoluted with $\Phi^{(0)}$, the partons entering the
next-to-leading (NLO) hard kernel $H^{(1)}$ are still on-shell. To
acquire a nontrivial $k_T$ dependence, $H^{(1)}$ must be
convoluted with the higher-order wave functions $\Phi^{(i)}$,
$i\ge 1$: the gluon exchanges in $\Phi^{(i)}$ render the incoming
partons of $H^{(1)}$, ie., the incoming partons of the quark
diagrams $G^{(1)}$ and the effective diagrams $\Phi^{(1)}$
off-shell by $k_T^2$ \cite{NL07}. We thus derive
$H^{(1)}(x,Q^2,k_T)$ according to the formula,
\begin{eqnarray}
& &H^{(1)}(x,Q^2,k_T)=G^{(1)}(x,Q^2,k_T)\label{pa1}\\
& &-\int dx' d^2k'_T
\Phi^{(1)}(x,k_T;x',k'_T)H^{(0)}(x',Q^2,k'_T)\;,\nonumber
\end{eqnarray}
with the $\bar q$ quark carrying the momentum $k=(xP_1^+,0,{\bf
k}_T)$. This is the way to obtain a $k_T$-dependent hard kernel
without breaking gauge invariance, since the gauge dependences
cancel between $G^{(1)}$ and $\Phi^{(1)}$. A physical quantity is
wtitten as a convolution of a hard kernel with model wave
functions, which are determined by methods beyond a perturbation
theory, such as lattice QCD and QCD sum rules, or extracted from
experimental data. A gauge-invariant hard kernel then leads to
gauge-invariant predictions from $k_T$ factorization.

\section{$B$ Decays in QCDF, SCET$_{\not 0}$, and PQCD}

Factorization theorems have been applied to exclusive $B$ meson
decays, and different approaches have been developed. Below we
compare the frameworks of perturbatiove QCD (PQCD)
\cite{LY1,KLS,LUY}, QCD-improved factorization (QCDF) \cite{BBNS},
and soft-collinear effective theory (SCET) \cite{bfl,bfps}. The
$B\to\pi$ transition form factor $F^{B\pi}$ involved in the
semileptonic decay $B\to\pi l\nu$ is expressed, in collinear
factorization, as
\begin{eqnarray}
F^{B\pi}=\int dx_1dx_2\phi_B(x_1)H(x_1,x_2) \phi_\pi(x_2),
\end{eqnarray}
with the LO hard kernel $H^{(0)}\propto (1+2x_2)/(x_1x_2^2)$. The
parton momentum fractions $x_1$ and $x_2$ are carried by the
spectator quarks
%momenta $k_1=(x_1P_1^+,0,{\bf 0}_T)$ and $k_2=x_2P_2$
on the $B$ meson and pion sides, respectively. Obviously, the
above integral is logarithmically divergent for the asymptotic
model $\phi_\pi\propto x(1-x)$ \cite{SHB}.

An end-point singularity implies that exclusive $B$ meson decays
are dominated by soft dynamics. That is, a heavy-to-light form
factor is not calculable in collinear factorization, and
$F^{B\pi}$ should be treated as a soft object \cite{BBNS}. This is
the basis of QCDF, and subleading corrections are added
systematically \cite{BF}. The above treatment has been further
elucidated in the framework of SCET \cite{BPS}: only the 1 term in
$H^{(0)}$ contains the end-point singularity, which leads to an
$O(\Lambda)$, ie., soft object $f^{\rm NF}$. The $2x_2$ term does
not, leading to an $O(\sqrt{m_B\Lambda})$ object $f^{\rm F}$ with
the $B$ meson mass $m_B$. Therefore, at leading power in $1/m_B$,
the $B\to \pi$ form factor can be split into the nonfactorizable
and factorizable components,
\begin{eqnarray}
F^{B\pi} &=& f^{\rm NF}+f^{\rm F}\;,
\end{eqnarray}
which have different power counting in the strong coupling
constant $\alpha_s$: the former is of $O(\alpha_s^0)$ and the
latter of $O(\alpha_s)$. The values of $f^{\rm NF}$ and $f^{\rm
F}$ have been determined from a fit to the $B\to\pi\pi$ data
\cite{BPRS}.

The formulation of the $B\to\pi$ transition in $k_T$ factorization
theorem is different. When the parton transverse momenta are
included, $f^{\rm NF}$ does not develop an end-point singularity,
and both $f^{\rm NF}$ and $f^{\rm F}$ are factorizable. Hence,
they are of the same order in $\alpha_s$, and can be combined into
a single term, giving \cite{LY1,TLS},
\begin{eqnarray}
F^{B\pi}&=&\int dx_1 dx_2 d^2k_{1T} d^2k_{2T}
\Phi_B(x_1,k_{1T})\nonumber\\
&\times& H(x_1,x_2,k_{1T},k_{2T}) \Phi_\pi(x_2,k_{2T})\;.
\end{eqnarray}
The end-point singularity is smeared into the large logarithm
$\ln^2(m_B/k_T)$, and absorbed into the pion wave function
$\Phi_\pi$. Resumming this logarithm to all orders in the
conjugate $b$ space \cite{BS,LS}, we derived the Sudakov factor
$S(m_B, b)$, which describes the parton distribution in $b$. Since
$f^{\rm NF}$ has been included, the large-energy symmetry
\cite{BF} is respected in PQCD. Recently, it was proposed that the
end-point singularity is attributed to a double counting of soft
degrees of freedom in collinear factorization \cite{MS06}. The
zero-bin subtraction removes the double counting, and leads to a
modified SCET formalism for $f^{\rm NF}$, labelled by SCET$_{\not
0}$ hereafter. The power counting of SCET$_{\not 0}$ in both
$1/m_B$ and $\alpha_s$ is then consistent with that of PQCD. The
regularization of the end-point singularity introduces the
logarithms $\ln\mu_\pm$ in SCET$_{\not 0}$ \cite{MS06}, whose
treatment has not yet been explained.

When applying the above factorization approaches to two-body
nonleptonic $B$ meson decays, further difference appears in the
treatment of annihilation amplitudes. The $O(\alpha_s m_0/m_B)$
annihilation amplitudes from the scalar penguin operators are
divergent in collinear factorization, where $m_0$ is the chiral
enhancement scale. Because of the end-point singularity, an
annihilation amplitude has been parameterized as
\begin{eqnarray}
\alpha_s\ln\frac{m_B}{\Lambda}\left(1+ \rho_{A}e^{i \delta_{A}}
\right)\;,
\end{eqnarray}
in QCDF \cite{BBNS}, where $\Lambda$ is a hadronic scale and the
free parameter $\rho_{A}$ is postulated to vary in the range $0\le
\rho_{A}\le 1$. It is not clear what mechanism generates the
strong phase $\delta_{A}$. With the similar zero-bin subtraction,
an annihilation amplitude is factorizable in SCET$_{\not 0}$, but
found to be almost real \cite{ALRW06}. A strong phase can only be
generated at loop level, ie., at $O(\alpha_s^2\Lambda/m_B)$.
However, we notice that the residual momentum carried by the $b$
quark in a nonfactorizable annihilation amplitude could result in
a strong phase of $O(\alpha_sm_0\Lambda/m_B^2)$ \cite{CLM07}, a
new mechanism not included in \cite{ALRW06}.

The scalar penguin annihilation amplitude is also factorizable in
$k_T$ factorization with the absence of the end-point singularity.
Furthermore, it was almost imaginary in PQCD \cite{KLS}, whose
corresponding mechanism is similar to the Bander-Silverman-Soni
one \cite{BSS}: when the $u$ or $c$ quark in a loop goes on mass
shell, a strong phase is produced. In the case of the annihilation
topology for heavy-to-light decays, the loop is formed by the
virtual particles in the LO PQCD hard kernel and the infinitely
many Sudakov gluons exchanged between two partons in a light
meson. The virtual particle acquires the transverse momentum $k_T$
through the Sudakov gluon exchange. A sizable strong phase is then
given, in terms of the principle-value prescription for the
virtual particle propagator, by
\begin{eqnarray}
\frac{1}{xm_B^2-k_T^2+i\epsilon}=\frac{P}{xm_B^2-k_T^2}
-i\pi\delta(xm_B^2-k_T^2).
\end{eqnarray}
Therefore, the treatment and the effect of the scalar penguin
annihilation amplitude are very different in QCDF, SCET$_{\not
0}$, and PQCD.

Though the scalar penguin annihilation amplitude is factorizable
in both PQCD and SCET$_{\not 0}$, it is almost imaginary in the
former, but real in the latter. We argue that the above different
opinions can be discriminated by comparing the direct CP
asymmetries in the charged $B$ meson decays $B^\pm\to K^\pm\pi^0$
and $B^\pm\to K^\pm\rho^0$. The $B^\pm\to K^\pm\pi^0$ decays
involve a $B$ meson transition to a pseudoscalar meson, so the
penguin emission amplitude is proportional to the constructive
combination of the Wilson coefficients $a_4+2(m_{0K}/m_B)a_6$,
where $m_{0K}$ is the chiral enhancement scale associated with the
kaon. The $B^\pm\to K^\pm\rho^0$ decays involve a $B$ meson
transition to a vector meson, so the penguin emission amplitude is
proportional to the destructive combination
$a_4-2(m_{0K}/m_B)a_6$. The annihilation effect is then less
influential in the former than in the latter. If the scalar
penguin annihilation is real, both decays will exhibit small
direct CP asymmetries, ie., $A_{CP}(B^\pm\to K^\pm\pi^0)\approx
A_{CP}(B^\pm\to K^\pm\rho^0)$. If the scalar penguin annihilation
is imaginary, it will cause a larger direct CP asymmetry in
$B^\pm\to K^\pm\rho^0$, ie., $A_{CP}(B^\pm\to K^\pm\pi^0)\ll
A_{CP}(B^\pm\to K^\pm\rho^0)$. The current data $A_{CP}(B^\pm\to
K^\pm\pi^0)=0.047\pm 0.026$ and $A_{CP}(B^\pm\to
K^\pm\rho^0)=0.31^{+0.11}_{-0.10}$ \cite{HFAG} seem to prefer an
imaginary scalar penguin annihilation.

\section{Recent Results}

\subsection{The $B\to\pi\pi$ Puzzle}

According to a naive estimate of the color-suppressed tree
amplitude, the hierarchy of the branching ratios
$B(B^0\to\pi^0\pi^0)\sim O(\lambda^2)B(B^0\to\pi^\mp\pi^\pm)$ is
expected. However, the data \cite{HFAG}
\begin{eqnarray}
B(B^0\to\pi^\mp\pi^\pm)&=&(5.2\pm 0.2)\times 10^{-6}\;,\nonumber\\
B(B^0\to\pi^0\pi^0)&=&(1.31\pm 0.21)\times 10^{-6}\;,\label{data}
\end{eqnarray}
show $B(B^0\to\pi^0\pi^0)\sim O(\lambda)B(B^0\to\pi^\mp\pi^\pm)$,
giving rise to the $B\to\pi\pi$ puzzle. As observed in
\cite{LM06}, the NLO corrections, despite of increasing the
color-suppressed tree amplitude significantly, are not enough to
enhance the $B^0\to\pi^0\pi^0$ branching ratio to the measured
value. A much larger color-suppressed tree amplitude, about the
same order as the color-allowed tree amplitude, must be obtained
in order to resolve the puzzle \cite{Charng2,P06}. To make sure
that the above NLO effects are reasonable, the PQCD formalism has
been applied to the $B\to\rho\rho$ decays \cite{LM06}, which also
receive the color-suppressed tree contribution. It was found that
the NLO PQCD predictions are in agreement with the data
$B(B^0\to\rho^0\rho^0)=(1.16\pm 0.46)\times 10^{-6}$ \cite{HFAG}.
We conclude that it is unlikely to accommodate the measured
$B^0\to\pi^0\pi^0$ and $\rho^0\rho^0$ branching ratios
simultaneously in PQCD, and that the $B\to\pi\pi$ puzzle remains.

It has been claimed that the $B\to\pi\pi$ puzzle is resolved in
the QCDF approach \cite{BBNS} with an input from SCET
\cite{BY05,BJ05,BJ052}: the inclusion of the NLO jet function, the
hard coefficient of SCET$_{\rm II}$, into the QCDF formula for the
color-suppressed tree amplitude gives sufficient enhancement of
the $B^0\to\pi^0\pi^0$ branching ratio, if adopting the parameter
scenario ''S4" \cite{BN}. It is necessary to investigate whether
the proposed new mechanism deteriorates the consistency of
theoretical results with other data. The formalism in \cite{BY05}
has been extended to the $B\to\rho\rho$ decays as a check
\cite{LM06}. It was found that the NLO jet function overshoots the
observed $B^0\to\rho^0\rho^0$ branching ratio very much as
adopting ''S4". That is, it is also unlikely to accommodate the
$B\to\pi\pi$ and $\rho\rho$ data simultaneously in QCDF.

\subsection{The $B\to\phi K^*$ Puzzle}

\begin{table}[ht]
\begin{center}
\caption{Polarization fractions in the penguin-dominated $B\to VV$
decays. }
\begin{tabular}{|c|c|c|}\hline
Mode&Belle&BaBar \\
\hline
$\phi K^{*+}$&$0.52\pm 0.08\pm 0.03$&$0.46\pm0.12\pm0.03$ \\
\hline$\phi K^{*0}$&$0.45\pm 0.05\pm 0.02$
&$0.52\pm0.05\pm0.02$\\
\hline$K^{*+}\rho^0$&&$0.9\pm0.2$ \cite{Bar}\\
\hline$K^{*0}\rho^+$&$0.43\pm 0.11^{+0.05}_{-0.02}$&$0.52\pm 0.10\pm0.04$\cite{Bar} \\
\hline
\end{tabular}
\label{tab:tab1}
\end{center}
\end{table}

For penguin-dominated $B\to VV$ decays, such as those listed in
Table~\ref{tab:tab1} \cite{HFAG}, the polarization fractions
deviate from the naive counting rules based on kinematics
\cite{CKL2}. This is the so-called the $B\to\phi K^*$ puzzle. Many
attempts to resolve the $B\to \phi K^*$ polarizations have been
made \cite{LM04}, which include new physics
\cite{G03,YWL,KDY,CG0504,HKW}, the annihilation contribution
\cite{AK,BRY06} in the QCDF approach, the charming penguin in SCET
\cite{BPRS}, the rescattering effect \cite{CDP,LLNS,CCS}, and the
$b\to sg$ (the magnetic penguin) \cite{HN} and $b\to s\gamma$
\cite{BRY} transitions. The annihilation contribution from the
scalar penguin operators improves the consistency with the data,
because it is of the same order for all the three final helicity
states, and could enhance the transverse polarization fractions
\cite{CKL2}. However, the PQCD analysis of the scalar penguin
annihilation amplitudes indicates that the $B\to\phi K^*$ puzzle
can not be resolved completely \cite{LM04}. A reduction of the
$B\to K^*$ form factor $A_0$, which is associated with the
longitudinal polarization, further helps accommodating the data
\cite{L04}. Note that there has not yet been any measurement,
which constrains $A_0$. Hence, the value of $A_0$ should have a
large uncertainty.

The penguin-dominated $B\to K^*\rho$ decays are expected to
exhibit similar polarization fractions. This is the reason the
longitudinal polarization fraction in the $B^+\to K^{*0}\rho^+$
decay, which contains only the penguin contribution, is close to
$f_L(\phi K^*)\sim 0.5$ as listed in Table~\ref{tab:tab1}. Another
mode $B^+\to K^{*+}\rho^0$, nevertheless, shows a large
longitudinal polarization fraction almost unity. This mode
involves tree amplitudes, which are subdominant, and should not
cause a significant deviation from $f_L\sim 0.5$. Though the data
of $f_L(K^{*0}\rho^+)$ from BaBar still suffer a large error
\cite{Bar}, the dramatically different longitudinal polarization
fractions, $f_L(K^{*+}\rho^0)\not=f_L(K^{*0}\rho^+)$, demand a
deeper understanding. It is highly suggested to update or perform
the measurement of $f_L(K^{*+}\rho^0)$. It is also worthwhile to
investigate the $B\to K^*K^*$ decays \cite{L04,DGLN}, whose
measurement can help discriminating the various proposals for
resolving the $B\to\phi K^*$ puzzle.

\subsection{The $B\to K\pi$ Puzzle}

The $B^0\to K^\pm\pi^\mp$ decays depend on the tree amplitude $T'$
and the QCD penguin amplitude $P'$. The data of the direct CP
asymmetry $A_{CP}(B^0\to K^\pm\pi^\mp)\approx -10\%$ then imply a
sizable relative strong phase between $T'$ and $P'$, which
verifies the LO PQCD prediction made years ago \cite{KLS}. The
$B^\pm\to K^\pm\pi^0$ decays contain the additional
color-suppressed tree amplitude $C'$ and electroweak penguin
amplitude $P_{ew}'$. Since both $C'$ and $P_{ew}'$ are
subdominant, the approximate equality for the direct CP
asymmetries $A_{CP}(B^\pm\to K^\pm\pi^0)\approx A_{CP}(B^0\to
K^\pm\pi^\mp)$ is expected. However, this naive expectation is in
conflict with the data \cite{HFAG},
\begin{eqnarray}
A_{CP}(B^0\to K^\pm\pi^\mp)&=&-0.093\pm 0.015\nonumber\\
A_{CP}(B^\pm\to K^\pm\pi^0)&=&0.047\pm 0.026\;,
\end{eqnarray}
leading to one of the $B\to K\pi$ puzzles.

While LO PQCD gives a negligible $C'$ \cite{KLS,LUY}, it is
possible that this supposedly tiny amplitude receives a
significant subleading correction. Note that the small $C'$ is
attributed to the accidental cancellation between the Wilson
coefficients $C_1$ and $C_2/N_c$ at the scale of the $b$ quark
mass $m_b$. In \cite{LMS05} the important NLO contributions to the
$B\to K\pi$ decays from the vertex corrections, the quark loops,
and the magnetic penguins were calculated. It was observed that
the vertex corrections increase $C'$ by a factor of 3, and induce
a large phase about $-80^o$ relative to $T'$. The large and
imaginary $C'$ then renders the total tree amplitude $T'+C'$ more
or less parallel to the total penguin amplitude $P'+P'_{ew}$ in
the $B^\pm\to K^\pm\pi^0$ decays, leading to nearly vanishing
$A_{CP}(B^\pm\to K^\pm\pi^0)=(-1^{+3}_{-6})\%$ at NLO (it is about
-8\% at LO). We conclude that the $B\to K\pi$ puzzle has been
alleviated, but not yet gone away completely. Whether new physics
effects \cite{BFRS,BL07} are called for will become clear when the
data get precise. More detailed discussion on this subject can be
found in \cite{G07}.

\subsection{Nonleptonic $B_s$ Decays}

Two-body nonleptonic $B_s$ meson decays are interesting, since
their study can test SU(3) or U-spin symmetry. The framework for
these decays is basically identical to that for $B_{u,d}$ meson
decays. The results of two-body nonleptonic $B_s$ meson decays
from different factorization approaches can be found in \cite{BN}
for QCDF, in \cite{WZ0610} for SCET, and in \cite{XCG0608,AKLL07}
for PQCD. Roughly speaking, the branching ratios predicted by QCDF
and by PQCD are similar, but the predicted direct CP asymmetries
are usually opposite in sign.

\section{Conclusion}

The factorization approaches are systematic theoretical tools for
exclusive $B$ meson decays, in which hadronic inputs are
universal, and the hard kernels can be computed order by order.
NLO corrections have been obtained for some $B$ meson decay modes,
and the consistency between the theoretical predictions and the
experimental data is improved in general. More need to be done in
order to pin down QCD uncertainty especially for those quantities
exhibiting puzzling behaviors. Higher-power corrections are
another important source of QCD uncertainty, which deserves a
careful investigation. The recent development in SCET$_{\not 0}$
is encouraging, whose counting rules become consistent with those
in PQCD. However, the arbitrary logarithms $\ln\mu_\pm$ resulting
from the zero-bin subtraction needs to be handled (recall that the
double logarithm $\ln^2 (m_B/k_T)$ from the smearing of the
end-point singularity has been resummed in PQCD). We did not
review the progress on the $\Delta S$ puzzle appearing in the
extraction of the weak phase $\sin(2\phi_1)$ from
penguin-dominated $B$ meson decays. For the detail, refer to
\cite{Z}.

%Because this template has been set up to meet
%requirements for conference proceedings  papers, it is important to maintain
%these established styles.
%Other editorial guidelines are described in the next section.

\begin{acknowledgments}
The work was supported in part by the National Science Council of
R.O.C. under Grant No. NSC-95-2112-M-050-MY3, and by the National
Center for Theoretical Sciences of R.O.C..

%This document is adapted from the instructions provided to the authors
%of the proceedings papers at FPCP~06, Vancouver, Canada~\cite{fpcp06},
%and from eConf templates~\cite{templates-ref}.
\end{acknowledgments}

\bigskip % extra skip inserted

\end{document}